\documentclass[graybox]{svmult}

\usepackage{mathptmx}       
\usepackage{helvet}         
\usepackage{courier}        
\usepackage{type1cm}        
\usepackage{makeidx}         
\usepackage{graphicx}        
\usepackage{multicol}        
\usepackage[bottom]{footmisc}

\makeindex             

\begin{document}

\title*{Primordial Black Holes and Quantum Effects}
\author{B. J. Carr}
\institute{B. J. Carr \at Department of Physics and Astronomy, Queen Mary University of London, 
Mile End Rd, 
London E1 4NS, UK,
\email{B.J.Carr@qmul.ac.uk}}
%
%
\maketitle

\abstract{
Primordial black holes (PBHs) are of special interest because of the crucial role of quantum effects in their formation and evaporation. This means that they provide a unique probe of the early universe, high-energy physics and quantum gravity.  We highlight some recent developments in the subject, including improved limits on the fraction of the Universe going into evaporating PBHs in the mass range $ 10^9-10^{17}\,\mathrm g $ 
and the possibility of using PBHs to probe a cosmological bounce. 
}

\section{Introduction}
\label{sec:1}

A comparison of the cosmological density at a time $t$ after the big bang with the density associated with a black hole of mass $M$
shows that 
PBHs should have of order the particle horizon mass,
$M_H(t) \approx 10^{15}(t/10^{-23}s)$g, at
formation. They could thus span an enormous mass range: from $10^{-5}$g for those formed at $10^{-43}$s to $10^5M_{\odot}$ for those formed  at 1~s.  
By contrast, black holes forming at the present epoch could never be smaller than about $1M_{\odot}$.
However, the high density of the early Universe is  not a {\it sufficient} condition for PBH formation. One either needs large-amplitude density fluctuations, possibly of inflationary origin, so that overdense regions can eventually stop expanding and recollapse,
or some sort of cosmological phase transition at which PBHs can form spontaneously (eg. via  the collapse of cosmic loops or 
the collisions of  bubbles of broken symmetry). All these formation mechanisms depend in some sense  on quantum effects and they are discussed in detail in Ref.~\cite{cksy} (henceforth CKSY).

The realization that PBHs might be small prompted Hawking
to study their quantum consequences This led to his famous 
discovery \cite{h74}
that black holes radiate thermally with a temperature
$T \approx 10^{-7}(M/M_{\odot})^{-1}$K and
evaporate on a timescale 
$ \tau(M) \approx 10^{64}(M/M_{\odot})^3$y.
Only black holes smaller
than $10^{15}$g would have evaporated by the present epoch and $10^{15}$g ones would be exploding today. 
Since the latter
would be producing photons with energy of order 100~MeV, the observational limits on
the $\gamma$-ray background intensity imply
that their density could not exceed $10^{-8}$ times the
critical density.
Nevertheless, this does not preclude PBHs playing other important cosmological roles. Indeed, their study provides a unique probe of four areas of physics:
gravitational collapse for $M>10^{15}g$,
high energy physics for $M\sim 10^{15}g$, the early Universe for $M<10^{15}g$ and
quantum gravity for $M\sim10^{-5}g$. 

\begin{figure}[b]
\sidecaption
\includegraphics[scale=.59]{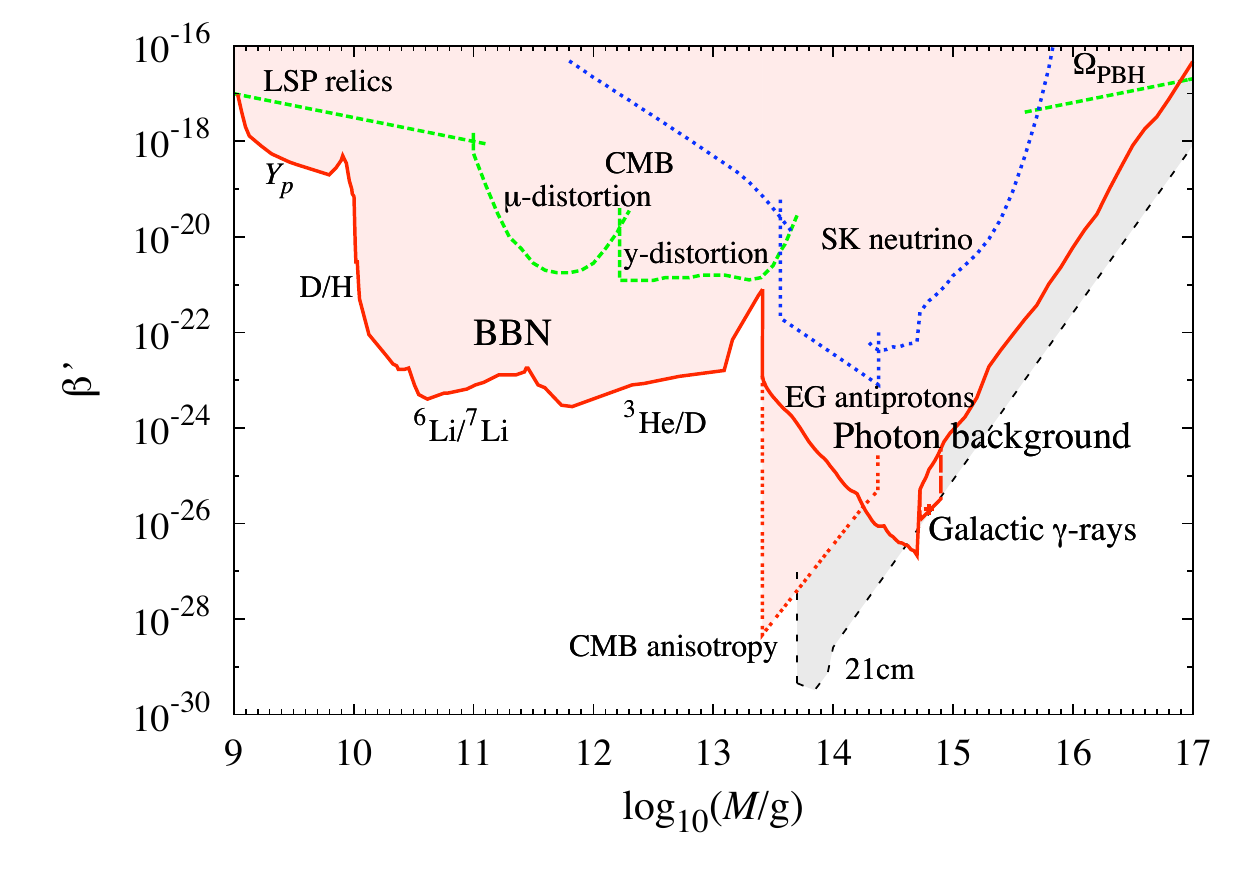}
\caption{
Combined BBN and EGB limits (solid), compared to other constraints on evaporating PBHs from LSP relics and CMB distortions (short-dashed), extragalactic antiprotons and neutrinos (dotted), the Galactic $ \gamma $-ray background (long-dashed), CMB anisotropies (dash-dotted)
and the density limit from the smallest unevaporated black holes (dashed). From Ref.~\cite{cksy}.
\label{fig:combined}
}
\end{figure}

Since both their formation and evaporation are a consequence of quantum effects, PBHs may offer the only astrophysical realization of what might be termed ``quantum black holes'' (i.e. holes for which quantum effects are important)  \cite{calmet}. 
This article will focus on their evaporation rather than their formation. In particular, it will discuss the upper limit on the fraction of the Universe going into PBHs as a function of mass  because this provides important constraints on models (such as inflation) predicting their formation. 
The fraction of the Universe collapsing into PBHs at time $t$ is related to their current density parameter $\Omega_{\rm PBH}$ by 
\begin{equation}
\beta 
\approx 10^{-6} \Omega_{\rm PBH} (t/s)^{1/2} 
\approx 10^{-18}\Omega_{\rm PBH}(M/10^{15}g)^{1/2}
\end{equation}
where the $t$ dependence reflects the decreasing ratio of the 
PBH and radiation densities at early times \cite{c75}.
Any limit on $\Omega_{\rm PBH}$ therefore places a constraint on $\beta$ as a function of $M$.
The constraints on $ \beta(M) $ have been studied by numerous authors
but the most recent and comprehensive discussion is that of Ref.~\cite{cksy}. The limits cover the mass range $ 10^9-10^{17}\,\mathrm g $ and are shown in Fig.~\ref{fig:combined}.
The important point is that the value of $ \beta(M) $ must be tiny throughout this mass range, so any cosmological model which entails an appreciable fraction of the Universe going into PBHs is immediately excluded.
The most stringent limits -- associated with big bang nucleosynthesis (BBN), the extragalactic $\gamma$-ray background (EGB) and observations of anisotropies  in the cosmic microwave background (CMB) --  
are discussed below. Positive evidence for PBHs might come from cosmic rays or short-period gamma-ray bursts but this is not covered below since the status of the observations is still ambiguous.

\section{
PBH Evaporations 
\label{sec:evap}
}

A black hole with mass $ M \equiv M_{10} \times 10^{10}\,\mathrm g $ emits thermal radiation with temperature
\begin{equation}
T_\mathrm{BH}
= \frac{1}{8\pi\,G\,M}
\approx
  1.06\,M_{10}^{-1}\,\mathrm{TeV}\,.
\end{equation}
The average energy of the emitted particles is $(4-6) \, kT_\mathrm{BH}$, depending on their spin. 
Charge and angular momentum are neglected because these will be lost through quantum emission on a shorter timescale. 
The mass loss rate can be expressed as
\begin{equation}
\frac{\mathrm dM_{10}}{\mathrm dt}
= -5.34 \times 10^{-5}\,f(M)\,M_{10}^{-2}\,\mathrm s^{-1}\,.
\end{equation}
Here $ f(M) $ is a measure of the number of emitted particle species, normalised to unity for a black hole with $ M \gg 10^{17}\,\mathrm g $, this emitting only particles which are (effectively) massless: photons, neutrinos  and gravitons.
Holes with $ 10^{15}\,\mathrm g < M < 10^{17}\,\mathrm g $ emit electrons, while those with $ 10^{14}\,\mathrm g < M < 10^{15}\,\mathrm g $ also emit muons, which subsequently decay into electrons and neutrinos.

Once $ M $ falls to around $ 10^{14}\,\mathrm g $, a black hole can also begin to emit hadrons.
However, hadrons are
composite particles, 
made up of quarks held together by gluons, so
for temperatures exceeding 
$ \Lambda_\mathrm{QCD} = 250-300\,\mathrm{MeV} $, one would expect the emission of quark and gluon jets rather than hadrons \cite{m91}. 
The jets would subsequently  fragment into 
hadrons but only after travelling a distance $ \Lambda_\mathrm{QCD}^{-1} \sim 10^{-13}\,\mathrm{cm} $,
which is much larger than the size of the hole. The QCD fragmentation has been calculated using the \textsc{PYTHIA} \cite{cksy} and 
 \textsc{HERWIG} \cite{mw} codes but with similar results.
Since there are many quark and gluon degrees of freedom, 
the value of $ f $ should roughly quadruple once the QCD temperature is reached.
If we sum up the contributions from all particles in the Standard Model up to $ 1\,\mathrm{TeV} $, 
this gives $ f(M) = 15.35 $ and a lifetime
\begin{equation}
\tau
\approx
  407\,
  \left(\frac{f(M)}{15.35}\right)^{-1}\,M_{10}^3\,\mathrm s\,.
\label{eq:tau}
\end{equation}
The critical mass for which $ \tau $ equals the age of the Universe ($t_0 \approx 13.7\,\mathrm{Gyr} $) is 
$M_*
\approx
  5.1 \times 10^{14}$g, corresponding to
$ f_* =1.9$ and $ T_\mathrm{BH}(M_*) = 21\,\mathrm{MeV} $.

\begin{figure}[b]
\sidecaption
\includegraphics[scale=.5]{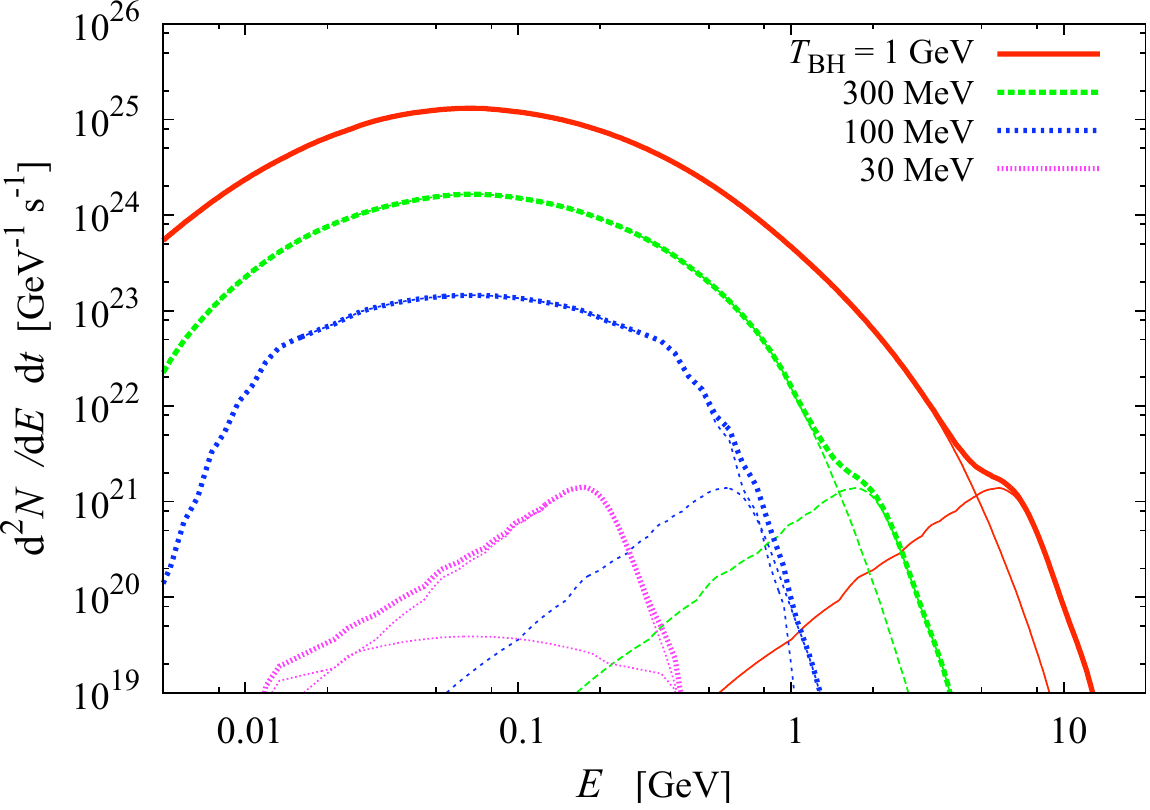}
\includegraphics[scale=.5]{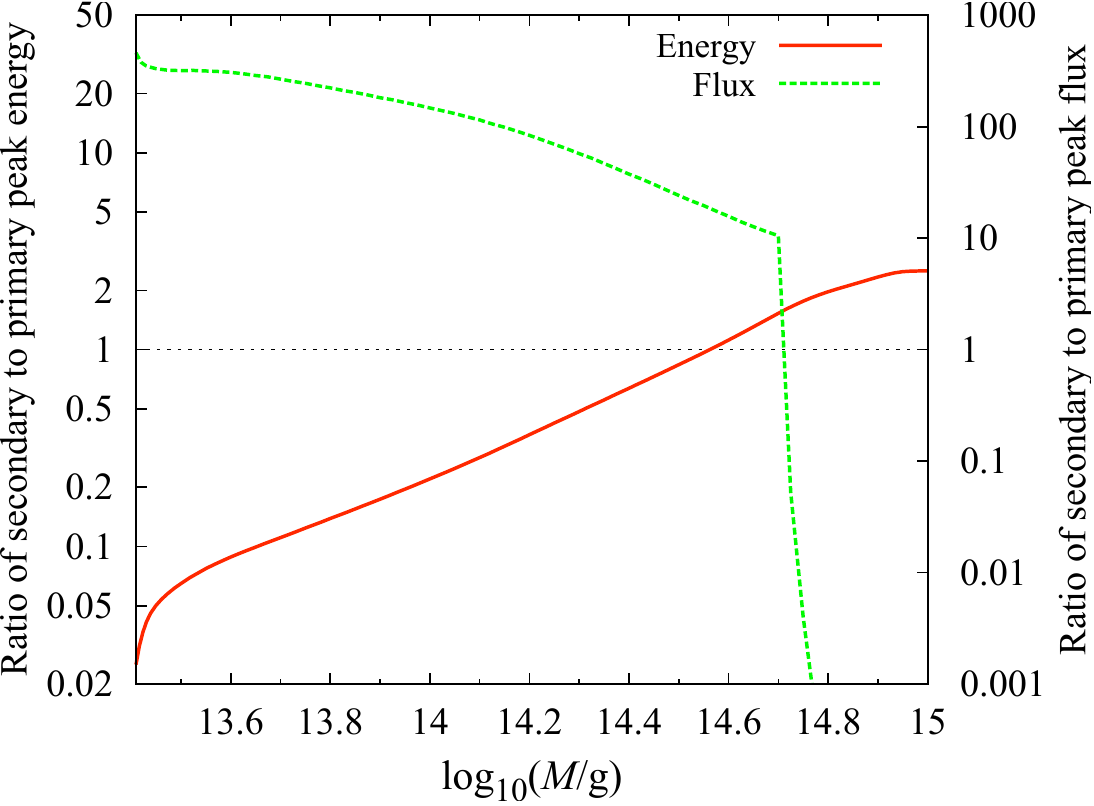}
\caption{
Left: Instantaneous emission rate of photons for four typical black hole temperatures, 
For each temperature, the curve with the peak to the right (left) represents the primary (secondary) component and the thick curve denotes their sum. Right: ratios of secondary to primary peak energies (solid) and fluxes (dashed). From Ref.~\cite{cksy}.
\label{fig:rate}
}
\end{figure}

The direct Hawking emission is termed the \emph{primary} component, while the jet fragmentation emission is termed the \emph{secondary} component.
The spectrum of secondary photons is dominated by the $ 2\gamma $-decay of soft neutral pions and peaks around $ E_\gamma \simeq m_{\pi^0}/2 \approx 68\,\mathrm{MeV} $.
The emission rates of primary and secondary photons for four typical temperatures are shown in Fig.~\ref{fig:rate}.
Although QCD effects are initially small for PBHs with $ M = M_* $, only contributing a few percent, they become important once $ M $ falls to 
$M_q \approx 0.4 M_* \approx 2 \times 10^{14}\,\mathrm g$
since the peak energy becomes comparable to $ \Lambda_\mathrm{QCD} $ then.
This means that an appreciable fraction of the time-integrated emission from the PBHs evaporating at the present epoch goes into quark and gluon jet products. However, a PBH with somewhat larger initial mass, $ M = (1+\mu)\,M_* $
will today have a mass
$M(t_0)
\approx
  (3\,\mu)^{1/3} M_*$ for $\mu \ll 1$.
Since this falls below $ M_\mathrm q $ only for
$ \mu < 0.02 $,
the fraction of the black hole mass going into secondaries falls off sharply above $ M_* $.
The ratio of the secondary to primary peak energies and the ratio of the time-integrated fluxes are shown as functions of $M$ in Fig.~\ref{fig:rate}.

There has been some dispute about the interactions between emitted particles 
beyond the QCD scale.
The usual assumption 
that there is no interaction has been refuted by Heckler \cite{hec1},
who claims that QED interactions could produce an optically thick photosphere once the black hole temperature exceeds $ T_\mathrm{BH} = 45\,\mathrm{GeV} $.
He has proposed that a similar effect may operate at an even lower temperature, $ T_\mathrm{BH} \approx 200\,\mathrm{MeV} $, due to QCD effects \cite{hec2}.
Variants of these models and their astrophysical implications have been studied by various authors.
However, MacGibbon \textit{et al.} \cite{mcp}
have reviewed all these models and identified a number of 
effects which invalidate them.
They conclude that emitted particles do not interact sufficiently to form a QED photosphere and that the conditions for QCD photosphere formation could only be temporarily satisfied (if at all) when the black hole temperature is of order $ \Lambda_\mathrm{QCD} $.

\section{
Constraints on $ \beta(M) $ Imposed by BBN, EGB and CMB
}

PBHs with $ M \sim 10^{10}\,\mathrm g $ and $ T_\mathrm{BH} \sim 1\,\mathrm{TeV} $ have a lifetime $ \tau \sim 10^3\,\mathrm s $ and therefore evaporate at the epoch of big bang nucleosynthesis (BBN). 
The effect of these evaporations on BBN has been a subject of long-standing interest and jet-produced hadrons are particularly important.  
Long-lived hadrons remain in the ambient medium long enough to leave an observable signature on BBN.
These effects were first discussed 
for the relatively low mass PBHs evaporating in the early stages of BBN \cite{ky}
but the analysis has now been extended by CKSY
to incorporate the effects of heavier PBHs evaporating after BBN. 
 
High energy particles emitted by PBHs modify the standard BBN scenario in three different ways:
(1) high energy mesons and antinucleons induce extra interconversion between background protons and neutrons even after the weak interaction has frozen out in the background Universe;
(2) high energy hadrons dissociate light elements synthesised in BBN, thereby reducing $ {}^4\mathrm{He} $ and increasing $ \mathrm D $, $ \mathrm T $, $ {}^3\mathrm{He} $, $ {}^6\mathrm{Li} $ and $ {}^7\mathrm{Li} $;
(3) high energy photons generated in the cascade further dissociate $ {}^4\mathrm{He} $.
The PBH constraints 
depend on
the initial baryon-to-photon ratio 
(allowing for PBH entropy production)
and the ratio of the PBH number density to the entropy density, $ Y_\mathrm{PBH} \equiv n_\mathrm{PBH}/s $, which
is related to the initial mass fraction by
$\beta
= 5.4 \times 10^{21}\,
  (\tau/1\,\mathrm s)^{1/2}\, Y_\mathrm{PBH}$.

\begin{figure}[b]
\sidecaption
\includegraphics[scale=.28]{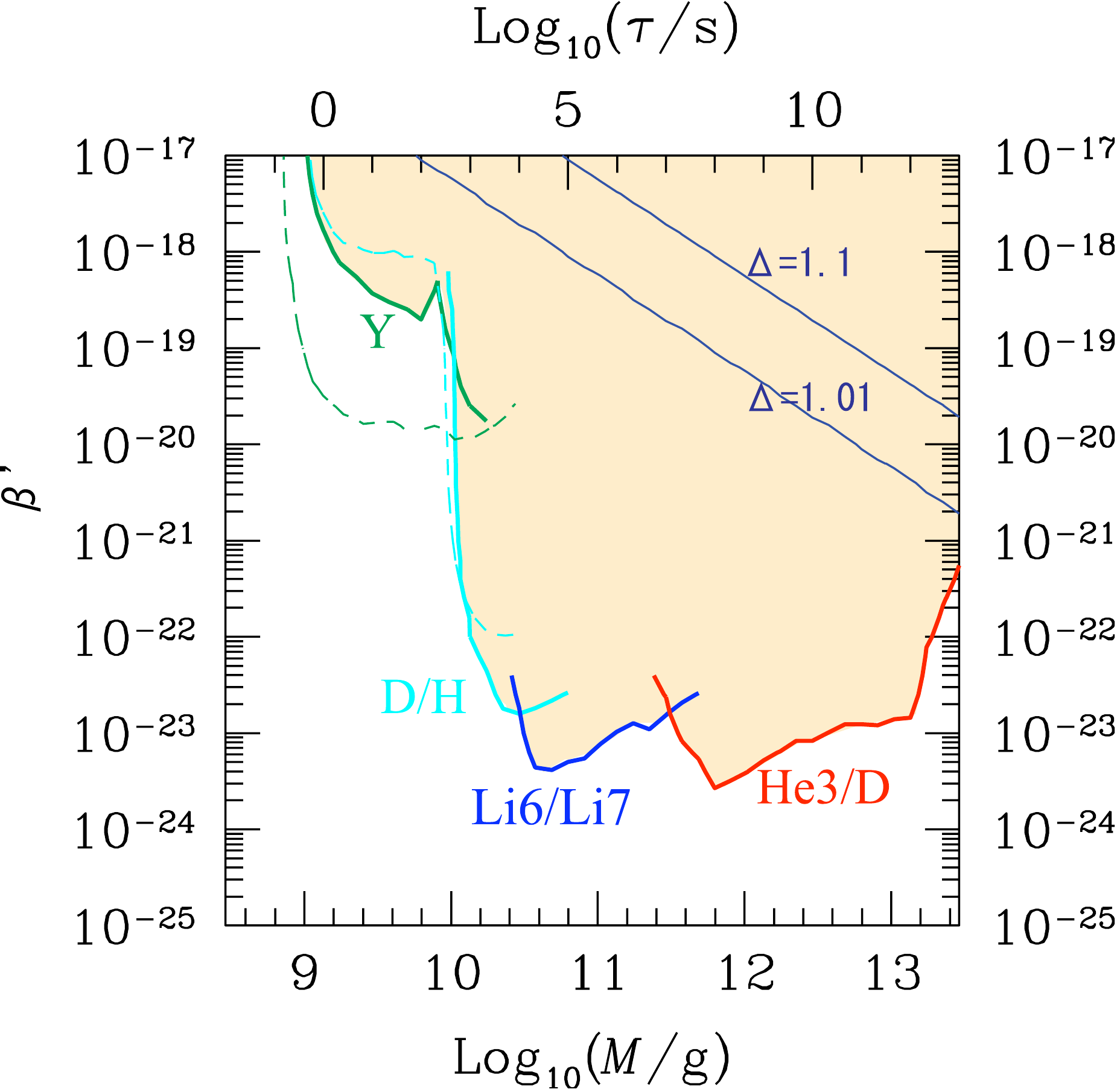}
\includegraphics[scale=.5]{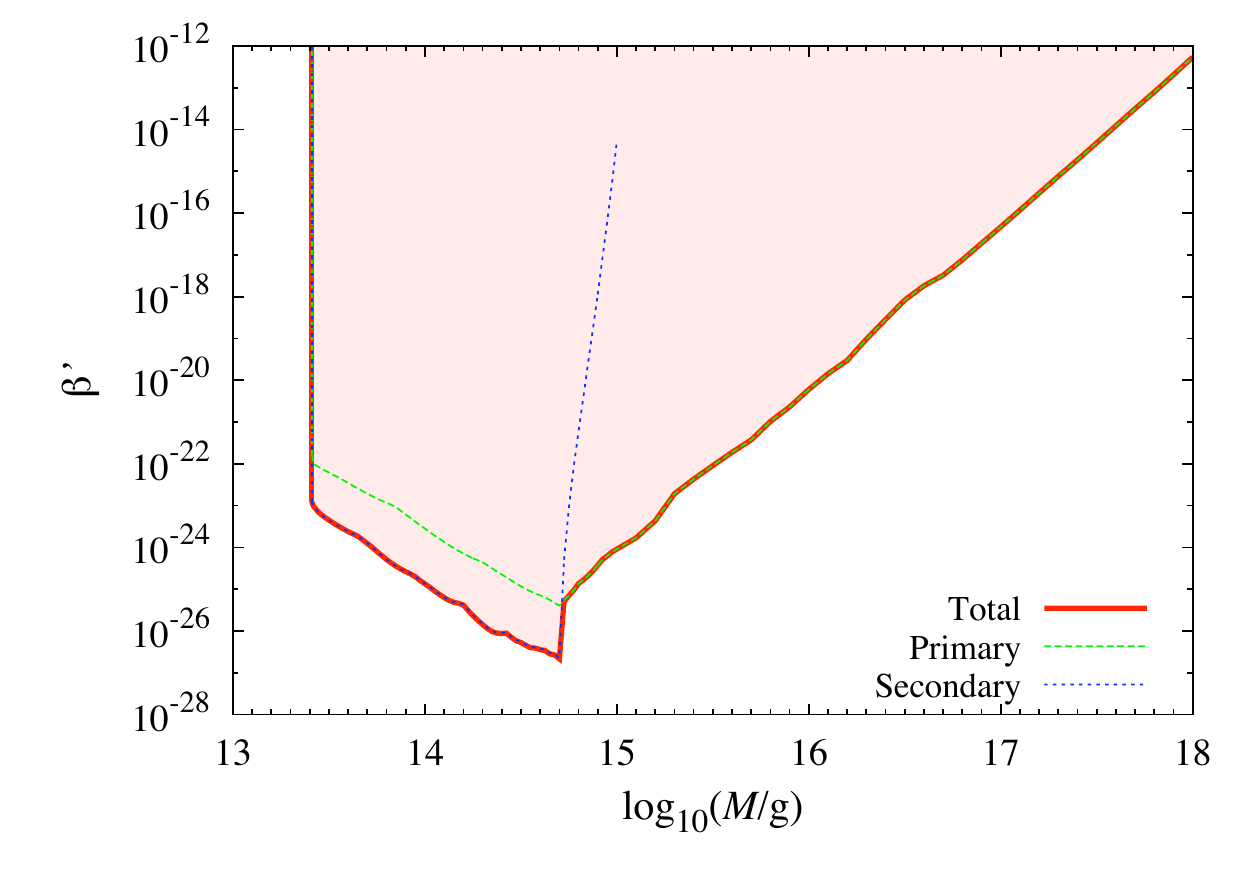}
\caption{
Left: Upper bounds on $ \beta(M) $ from BBN, with broken line giving earlier limit.
Right: Upper bounds on $ \beta(M) $ from the extragalactic photon background, with no other contributors to the background having been subtracted. From Ref.~\cite{cksy}. 
\label{fig:bbn}
}
\end{figure}
The results of these calculations are summarized in Fig.~\ref{fig:bbn}.
PBHs with $ M <  10^9\,\mathrm g $  ($ \tau < 10^{-2}$s) are free from BBN constraints because they evaporate before weak freeze-out.
PBHs with $ M = 10^9-10^{10}\,\mathrm g $ ($ \tau = 10^{-2}-10^2\,\mathrm s $) are constrained by process (1), those with $ M = 10^{10}-10^{12}\,\mathrm g $ ($ \tau = 10^2-10^7\,\mathrm s $) by process (2) and those with $ M > 10^{12}-10^{13} \,\mathrm g $ ($ \tau = 10^7-10^{12}\,\mathrm s $) by process (3). 
We also show as a broken line the limits obtained earlier \cite{ky}.
The helium limit is weaker because the primordial abundance is now known to be smaller, while the deuterium limit is stronger
because of its extra production by hadrodissociation of helium.

It has been known for $40$ years that observations of the diffuse extragalactic $\gamma$-ray background (EGB) 
constrain
$ \Omega_\mathrm{PBH} (M_*)$  to be less than 
around $ 10^{-8} $ \cite{ph}.
This limit has subsequently been refined by numerous authors and most recently by CKSY. 
In order to determine the present background spectrum of photons generated by PBH evaporations, one must integrate over the lifetime of the black holes,
allowing for the fact that particles generated in earlier cosmological epochs will be redshifted in energy by now.
The highest energy photons are associated with PBHs of mass $ M_* $.
Those from PBHs with $ M > M_* $ are at lower energies because they are cooler, while those from PBHs with $ M < M_* $ (although initially hotter) are at lower energies because they are redshifted.

The most recent X-ray and $ \gamma $-ray observations are summarized by CKSY
and correspond to an intensity $ I^\mathrm{obs} \propto E_{\gamma 0}^{-(1+\epsilon)} $ where $ \epsilon $ 
lies between $ 0.1 $ 
and $ 0.4 $.
The origin of these backgrounds is thought to be primarily distant astrophysical sources, such as blazars, and in principle one should remove these contributions
 before calculating the PBH constraints~\cite{barrau}.
CKSY do not attempt such a subtraction, so their constraints may be overly conservative. 
The limits on $ \beta(M)$ are shown in Fig.~\ref{fig:bbn} and depend on the relative magnitude of the primary and secondary components. 
PBHs with $ M > M_* $ can never emit secondary photons and one obtains 
$\beta(M)
\le
  4 \times 10^{-26}\,
  (M/M_*)^{7/2+\epsilon}$. Those with $ M \le M_* $ will do so once 
 $ M $ falls below $M_q \approx 2 \times 10^{14}\,\mathrm g $ and one obtains $\beta(M)
\le
  3 \times 10^{-27}\,
  (M/M_*)^{-5/2-2\,\epsilon}$.
These $ M $-dependences  explain the qualitative features of Fig.~\ref{fig:bbn}
and the associated limit on the density parameter is $ \Omega_\mathrm{PBH}(M_*) \le 5 \times 10^{-10} $.
Since photons emitted at sufficiently early times cannot propagate freely, there is a minimum mass $ M_\mathrm{min} \approx 3 \times 10^{13}$g below which the above constraint is inapplicable..

If PBHs of mass $ M_* $ are clustered inside our own Galactic halo, as expected, then there should also be a Galactic $ \gamma $-ray background.
Some time ago 
it was claimed that such a background had been detected by EGRET between $ 30\,\mathrm{MeV} $ and $ 120\,\mathrm{GeV}$
and that this could be attributed to PBHs \cite{w96}.
A more recent analysis of EGRET data between $ 70\,\mathrm{MeV} $ and $ 150\,\mathrm{GeV} $
gives a limit
$ \Omega_\mathrm{PBH}(M_*) \le 2.6 \times 10^{-9} $ or
$ \beta(M_*) < 1.4  \times 10^{-26} $  \cite{Lehoucq:2009ge}, which is a factor of 5 above the EGB constraint.  
However, CKSY point out that the EGB constraint on $\beta(M)$ comes from the time-integrated contribution of the $ M_* $ black holes, which peaks  at $120$~MeV,  
whereas the Galactic background is dominated by PBHs which are slightly larger than this.
The emission from PBHs with initial mass $ (1+\mu)\,M_* $ currently peaks at an energy $ E \approx 100\,(3\,\mu)^{-1/3}\,\mathrm{MeV} $, which is in the range $ 70$~MeV$-150$~GeV for $ 0.7 > \mu > 0.08 $. The corrected limit is shown in Fig.~\ref{fig:combined}.

\if
Photons from PBHs in the range $ 10^{11}\,\mathrm g < M < 10^{13}\,\mathrm g $, although partially thermalised, will produce noticeable distortions in the CMB spectrum.
Those emitted after the freeze-out of double-Compton scattering ($ t  > 7 \times 10^6\,\mathrm s $), corresponding to $ M > 10^{11} \mathrm g $, induce a $ \mu $-distortion,
while those emitted after the freeze-out of the single-Compton scattering ($ t > 3 \times 10^9\,\mathrm s $), corresponding to $M > 10^{12} \mathrm g $, induce a $ y $-distortion.
\fi
The CMB anisotropy constraint arises because electrons and positrons from PBHs heat the matter  content of the Universe after recombination, thereby damping small-scale anisotropies.
CKSY find
$\beta(M)
< 3 \times 10^{-30}\,
 (M/10^{13}\mathrm g)^{3.1}$
for $2.5 \times 10^{13}\,\mathrm g < M < 2.4 \times 10^{14}\,\mathrm g$.
The upper limit
corresponds to evaporation 
at the epoch of reionization ($z=6$), since
 the opacity is too low for emitted particles to heat the matter thereafter.
This is stronger than all the other limits in this mass range.

 \section{PBHs and Dark Matter}

Roughly $30\%$ of the total density of the Universe is now thought to be in the form of ``cold dark matter''. There has been a lot of interest in whether PBHs could provide this, since those larger than $10^{15}$g would not have evaporated yet and would certainly be massive enough to be dynamically cold.  
One possibility is that  PBHs with a mass of around $1M_{\odot}$ could have 
formed efficiently at the quark-hadron phase transition at $10^{-5}$s because of a temporary reduction in pressure \cite{j97}.
 At one stage there seemed to be evidence for this from microlensing observations.
The data no longer support this but there are no constraints excluding PBHs in the sublunar range $ 10^{20}\,\mathrm g < M < 10^{26}\,\mathrm g $ \cite{bbkp} or intermediate mass range $10^2\,M_\odot < M < 10^4\,M_\odot $ \cite{Saito}
from having an appreciable density. 

Some people have speculated that black hole evaporation could cease once the hole gets close to the Planck mass ($M_P$) due to the influence of extra dimensions, higher order corrections to the gravitational Lagrangian, string effects, the
 Generalized Uncertainty Principle etc. The resulting stable relics would then be natural candidates for the dark matter \cite{m87}.
In an inflationary scenario, if the relics have a mass $ \kappa\,M_\mathrm{Pl} $
and reheating occurs at a temperature $ T_\mathrm R $ (when the PBHs form), then the requirement that the relic density be less than the dark matter density 
implies 
$\beta(M)
< 2 \times 10^{-28}\,\kappa^{-1}\,(M/M_\mathrm{Pl})^{3/2}$
for $(T_\mathrm R/T_\mathrm{Pl})^{-2}
< M/M_\mathrm{Pl}
< 10^{11}\,\kappa^{2/5}$ \cite{cgl}.
The lower mass limit arises because PBHs generated before reheating are diluted exponentially. (If there is no inflationary period,  the constraint extends all the way down to the Planck mass.)
The upper mass limit arises because PBHs larger than this dominate the total density before they evaporate, in which case the current cosmological photon-to-baryon ratio is determined by the baryon asymmetry associated with their emission. 

\section{PBHs as a Probe of a Cosmological Bounce}

In some cosmological scenarios, the Universe is expected to eventually recollapse to a big crunch and then bounce into a new expansion phase. Such a bounce may arise through either classical or quantum gravitational effects.
Even if the universe is destined to expand forever, it may have been preceded by an earlier collapsing phase. 
Both past and future bounces would arise in cyclic models, as reviewed in Ref.~\cite{cc}.
It is therefore interesting to ask whether black holes could either be generated by a big crunch 
or survive it if they were formed earlier \cite{cc}. We refer to these as  
``big-crunch black holes'' (BCBHs) and ``pre-crunch black holes'' (PCBHs), respectively. 
If such black holes were detectable today, they would provide a unique probe of the last cosmological bounce, although this raises the question of whether one could 
differentiate between black holes formed just before and just after the last bounce.

\begin{figure}[b]
\sidecaption
 \includegraphics[scale=.37]{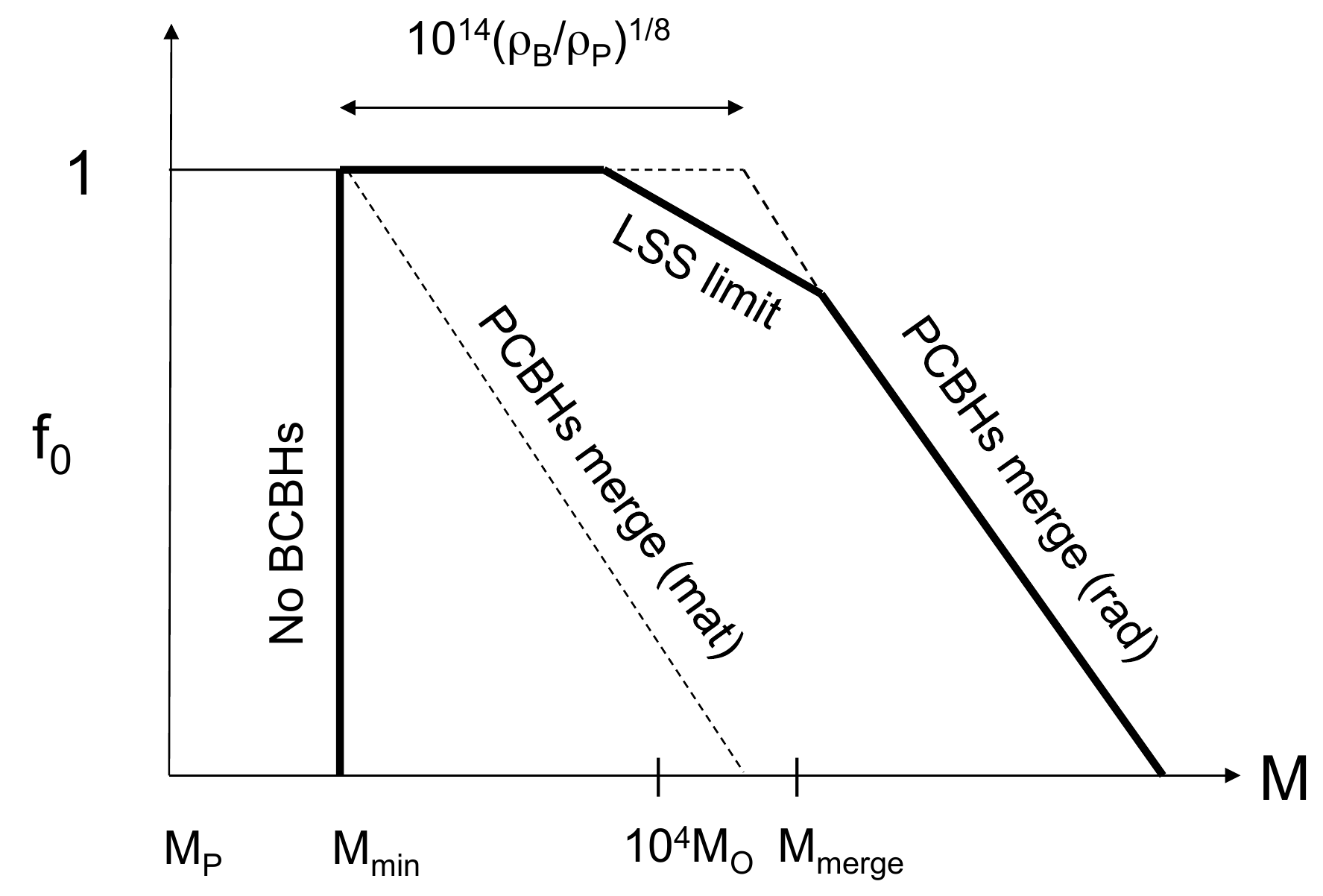}
\caption{ \label{bounce} This shows the domain in which black holes of mass $M$ containing a fraction $f_0$ of the present density can form in a big crunch or avoid merging if they exist before it. From Ref.~\cite{cc}.}
   \end{figure}

Let us assume that
the universe bounces at some density $\rho_B$, 
Since the density associated with a black hole of mass $M$ is $\rho_{BH} = (3M/4\pi R_{S}^3)$, 
this corresponds to a {\it lower} limit on the BCBH mass $M_{\mathrm{min}} \sim (\rho_P/\rho_B)^{1/2}M_P$.
There is also a mass range in which pre-existing PCBHs 
lose their individual identity by merging with each other prior to the bounce.  
If the fraction of the cosmological density in these black holes at the 
bounce epoch is $f_B$, 
then the average separation between them is less than their size (i.e. the black holes merge) for 
$M > f_B^{-1/2} M_{\mathrm{min}} $. The important point, as indicated in Fig.~\ref{bounce}, is that
there is a always range of masses in which BCBHs may form and PCBHs do not merge. 
However, one must distinguish between $f_B$ and the {\it present} fraction $f_0$ of the Universe's mass in black holes.   
Since the ratio of the black hole to radiation density 
scales as the cosmic scale factor, the fraction of the universe in black holes 
at a radiation-dominated bounce is
$f_B 
 \approx  
f_0 \left({\rho_\mathrm{eq}}/{\rho_B} \right) ^{1/4}$ where $\rho_{\mathrm{eq}} \sim 10^{12} \rho_0 \sim 10^{-17} \mathrm{g \, cm}^{-3}$. 
The merger condition therefore becomes $f_0 > 
10^{28}\left({\rho_\mathrm{B}}/{\rho_{\mathrm{P}}} \right) ^{-3/4}\left(M/M_P \right)^{-2} $, 
as indicated by the line on the right of Fig.~\ref{bounce}. 

There are various dynamical 
constraints on the form of the function $f_0(M)$ for non-evaporating PCBHs.
They must have $f_0 < 1$ in order not to exceed the
observed cosmological density and this gives a minimum  value for the merger mass,
$M_{\mathrm{merge}} \sim 10^{9} (t_B/ t_P) ^{3/4}$~g, where $t_B$ is the time of the bounce
as measured from the notional time of infinite density. 
This is around $10^{15}$~g for $t_B \sim 10^{-35}$~s but as large as 
$10^4 M_{\odot}$ for $t_B \sim 10^{-5}$s, so the observational consequences would be very significant. 
Another important constraint, deriving from Poisson fluctuations in the
black hole number density, is associated with
large-scale structure (LSS) formation~\cite{ams}.
This gives a limit $f_0 < (M/10^4 M_{\odot})^{-1}$, 
as shown by the line at the  top right of Fig.~\ref{bounce}.

\end{document}